\documentclass[preprint,showpacs,preprintnumbers,amsmath,amssymb,floatfix]{revtex4}
\usepackage{graphicx}
\begin{document}
\title{Analysis of Anomalous Interactions With Heavy Leptons
at $ep$, $e^+e^-$ and $pp$ Colliders}
\author{A. T. Tasci}
\email{tasci_a@ibu.edu.tr}
\author{A. T. Alan}
\email{alan_a@ibu.edu.tr} \affiliation{Abant Izzet Baysal
University, Department of Physics, 14280 Bolu, Turkey}
\begin{abstract}
We consider possible production of heavy leptons via anomalous
interactions at future $ep$ colliders (THERA and LHeC), $e^+e^-$
colliders (ILC and CLIC) and $pp$ collider CERN LHC. The
production, backgrounds and signatures of heavy leptons are
analyzed. We obtain the upper mass limits of 800 GeV at LHeC, 450
GeV at ILC and 650 GeV at LHC for optimal choices of relevant
parameters.
\end{abstract}
\maketitle
\section{Introduction}
Many models extending the standard theory of quarks and leptons
predict the existence of new generations of fermions. New heavy
leptons play an important role in the search for extensions of the
Standard Model (SM) and any signal for the production of such
fermions will play a milestone role in the discovery of new
physics. Many analysis have been done for the production of these
particles at future $e^-e^+$ \cite{Almeida:1990ay, Almeida:1994ad,
Almeida:2000yx, Almeida:2003cy}, at hadron \cite{Frampton:1992ik,
Coutinho:1998bu, CiezaMontalvo:2002gk, CiezaMontalvo:2000ys} and
also at $ep$ colliders \cite{Almeida:1990vt, Rizzo:1986wf,
Alan:2004rv, Alan:2006ux}. The experimental upper bounds for the
heavy lepton masses were found to be 44 GeV by OPAL
\cite{Akrawy:1990dx}, 46 GeV by ALEPH \cite{Decamp:1991uy} and 90
GeV by H1 \cite{Ahmed:1994yi} Collaborations. Hence, these leptons
could be detected at future high energy colliders as their masses
could be as high as 1 TeV. In this work, we analyze possible
production and decay processes of new heavy leptons via some
anomalous interactions in $ep$, $e^-e^+$ and $pp$ collisions at
future $ep$ colliders THERA \cite{Abramowicz:DESY-01-011} and LHeC
\cite{Dainton:2006wd}, at linear colliders ILC \cite{lc} and CLIC
\cite{Accomando:2004sz}, at the CERN Large Hadron Collider (LHC)
\cite{unknown:1995mi}. The main parameters of these colliders are
given in Table~\ref{tab:table1}.

In the SM, Flavor Changing Neutral Current (FCNC) processes
receive contributions from only higher order corrections. Here we
offer the following effective Lagrangian having magnetic moment
type operators to describe the interactions of the heavy leptons
by ordinary ones;
\begin{eqnarray}\label{e1}
    \mathcal{L}_{\mathtt{eff}}&=&\frac{ie\kappa_\gamma}{\Lambda}L\sigma_{\mu\nu}q^\nu l A^\mu
+\frac{g}{2\cos\theta_W}L\left[\gamma_\mu(c_v-c_a\gamma_5)+\frac{i\sigma_{\mu\nu}q^\nu}{\Lambda}\kappa_Z\right]lZ^\mu+h.c.,
\end{eqnarray}
where $\kappa_\gamma$ and $\kappa_Z$ are the anomalous magnetic
dipole moment factors, $c_v$ and $c_a$ are the corresponding
anomalous non-diagonal $Z$ couplings which are zero in the SM, $q$
is the momentum of the exchanged gauge boson, $\theta_W$ is the
Weinberg angle, $e$ and $g$ denote the gauge couplings relative to
$U(1)$ and $SU(2)$ symmetries respectively, $A^{\mu}$ and
$Z^{\mu}$ the fields of the photon and Z boson and $\Lambda$ is
the new physics scale. In the numerical calculations we have taken
$c_v$ and $c_a$ at the order of SM values.

Heavy leptons decay through the neutral current processes
$L\rightarrow \gamma l$ and $L\rightarrow Zl$ via anomalous
couplings in Eq.~(\ref{e1}), where $l$ denotes one of the ordinary
charged leptons ($e, \mu, \tau$). The branching ratios (BR) for
these processes would be around $33\%$ for each channel.
Neglecting ordinary lepton masses the decay widths are obtained
as,
\begin{eqnarray}\label{e2}
\Gamma(L\rightarrow
l\gamma)=\frac{\alpha\kappa_\gamma^2m_L^3}{2\Lambda^2},
\end{eqnarray}
\begin{eqnarray}\label{e3}
\Gamma(L\rightarrow
lZ)&=&\frac{\alpha(m_L^2-M_Z^2)^2}{16\Lambda^2m_L^3M_Z^2\sin^2\theta_W\cos^2\theta_W}
\bigg[\kappa_Z^2M_Z^4\nonumber\\&+&2M_Z^2\bigg[\Lambda^2(c_v^2+c_a^2)+(\kappa_Z^2m_L^2
-3c_v\kappa_Z\Lambda m_L)\bigg]+\Lambda^2(c_v^2+c_a^2)m_L^2
\bigg],
\end{eqnarray}
where $\alpha$ is the electromagnetic coupling constant, $M_Z$ and
$m_L$ refer to masses of $Z$ boson and decaying lepton,
respectively.

The only background processes related to the signal reaction
$ep\rightarrow LqX$ are $eq\rightarrow e\gamma q$ and
$eq\rightarrow eZq$. In Table~\ref{tab:table2} we have presented
the decay widths $\Gamma(L\rightarrow \gamma e)$ and
$\Gamma(L\rightarrow Ze)$ for a wide range of $m_L$. As seen from
this table the branching ratio $\mathrm{BR}(L\rightarrow\gamma e)$
is extremely small compared with $\mathrm{BR}(L\rightarrow Ze)$,
hence the background from $\gamma e$ channel can be safely
ignored.
\section{Production of Heavy Leptons in $ep$ Collisions}
Using Eq.~(\ref{e1}), the differential cross section for the
subprocess $eq\rightarrow Lq$, through the $t$ channel mediated by
$\gamma$ and $Z$ is obtained as;
\begin{eqnarray}\label{e4}
&&\frac{d\hat{\sigma}(eq\rightarrow
Lq)}{d\hat{t}}=\frac{2\kappa_\gamma^2e_q^2\pi\alpha^2}{\Lambda^2\hat{s}^2\hat{t}}
\bigg\{(2\hat{s}+\hat{t})m_L^2-2\hat{s}(\hat{s}+\hat{t})-m_L^4\bigg\}
\nonumber\\
&&+\frac{\pi\alpha^2}{8\Lambda^2\hat{s}^2\sin^4\theta_W\cos^4\theta_W
 [(\hat{t}-M_Z^2)^2
+M_Z^2\Gamma_Z^2]}\bigg\{2\kappa_Z\Lambda c_v^l({c_a^q}^2
+{c_v^q}^2)(m_L^2-\hat{t})m_L\hat{t}\nonumber\\
&&+4\Lambda c_v^qc_a^qc_a^l(\Lambda c_v^l-m_L\kappa_Z)(m_L^2
-2\hat{s}-\hat{t})\hat{t}\nonumber\\
&&
-\Lambda^2({c_a^q}^2+{c_v^q}^2)({c_a^l}^2+{c_v^l}^2)\left[(2\hat{s}
+\hat{t})m_L^2-2\hat{s}^2-2\hat{s}\hat{t}-\hat{t}^2\right]\nonumber\\
&&-\kappa^2_Z({c_a^q}^2+{c_v^q}^2)\left[m_L^4-(2\hat{s}+\hat{t})m_L^2
+2\hat{s}(\hat{s}+\hat{t})\right]\hat{t}\bigg\}\nonumber\\
&&+\frac{\kappa_\gamma
e_q\pi\alpha^2(\hat{t}-M_Z^2)}{\Lambda^2\hat{s}^2\sin^2
\theta_W\cos^2\theta_W
\left[(\hat{t}-M_Z^2)^2+M_Z^2\Gamma_Z^2\right]}
\bigg\{\kappa_Zc_v^q\left(m_L^4-(2\hat{s}+\hat{t})m_L^2+2\hat{s}
(\hat{s}+\hat{t})\right)\nonumber\\
&&+\Lambda
m_L\left(c_a^lc_a^q(m_L^2-2\hat{s}-\hat{t})+c_v^lc_v^q(\hat{t}
-m_L^2)\right)\bigg\},
\end{eqnarray}
where $e_q$ is quark charge in units of $e$, $\Gamma_Z$ is the
decay width of mediator $Z$. The total production cross section is
obtained by the integration of differential cross section over the
parton distributions in the proton as;
\begin{eqnarray}\label{e7}
\sigma(ep\rightarrow
LqX)=\int_{x_{min}}^{1}dxf_{q}(x,Q^2)\int_{t_{min}}^{t_{max}}\frac{d\hat{\sigma}}{d\hat{t}}d\hat{t},
\end{eqnarray}
where $x_{min}=m_{L}^{2}/s$, $\hat{t}_{min}=-(\hat{s}-m_{L}^{2})$
and $\hat{t}_{max}=0$. For the parton distribution functions
$f_{q}(x,Q^2)$ we have used the CTEQ5 parametrization
\cite{Lai:1999wy} providing the dependence on momentum transfer
which have been taken as $Q=m_L$ and for illustration, values of
$\kappa_\gamma=\kappa_Z=0.02$ have been taken for anomalous
magnetic moment couplings.

\begin{figure*}
\includegraphics[width=6cm, angle=-90]{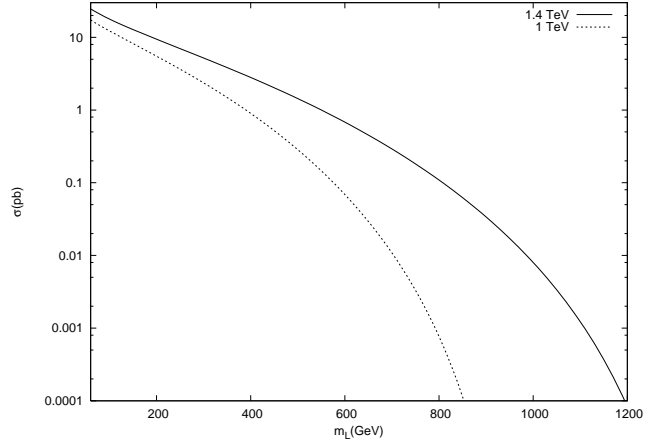}
\caption{The total cross sections as function of the heavy lepton
masses for lepton-hadron colliders THERA ($\sqrt{s}=$1 TeV) and
LHeC ($\sqrt{s}=$1.4 TeV).}\label{fig:f1}
\end{figure*}

\begin{figure*}
\includegraphics[width=6cm, angle=-90]{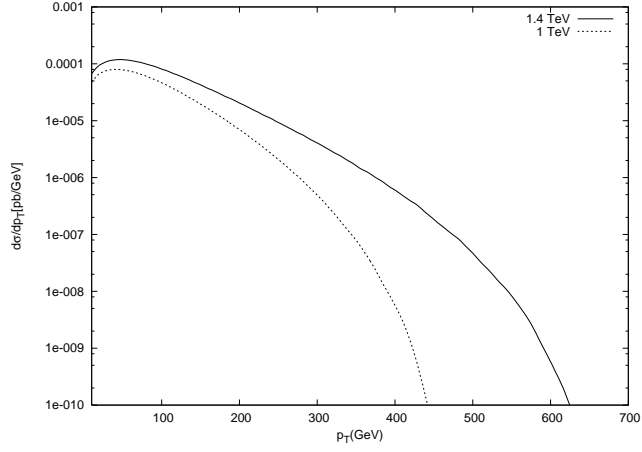}
\caption{$p_T$ distributions of the background for lepton-hadron
colliders THERA and LHeC.}\label{fig:f2}
\end{figure*}

\begin{figure*}
\includegraphics[width=6cm, angle=-90]{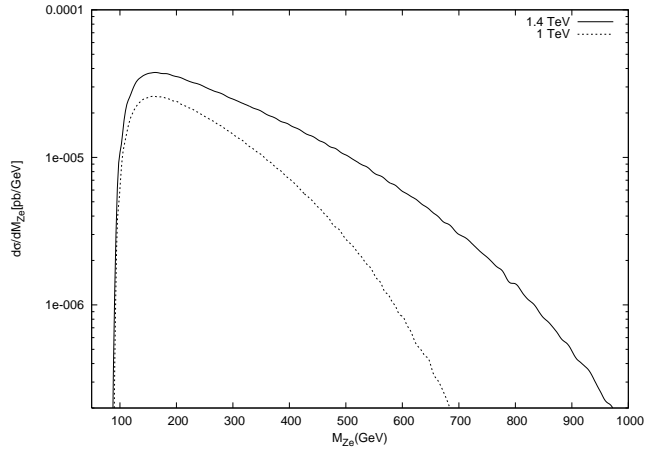}
\caption{The invariant mass distributions of the $Ze$ system for
the background for lepton-hadron colliders THERA and
LHeC.}\label{fig:f3}
\end{figure*}

We give the production cross sections for the signal as functions
of the heavy lepton masses, $m_L$, in Fig.~\ref{fig:f1} for
$\sqrt{s}$=1 and 1.4 TeV. We applied an optimal cut of
$|M_{Ze}-m_L|<50$ GeV for the considered heavy lepton mass range,
in order to contain all signal events that are smeared by the
experimental resolution. Fig.~\ref{fig:f2} shows the $p_T$
distributions of the final state particles for the relevant
background process $eq\rightarrow eZq$ in $ep$ collisions. These
$p_T$ distributions have peaks around 50 GeV and suppressed at
higher values. Fig.~\ref{fig:f3} shows the invariant mass
distributions of $Ze$ system, after a cut of $p_T^{e,j}>50$ GeV.

In Tables~\ref{tab:tab3} and \ref{tab:table3} we present the
production and total background cross sections depending on the
heavy lepton masses for suitable ranges at THERA and LHeC. The
number of signal events per year can easily be estimated from the
integrated luminosities of the colliders. The branching ratios
$\mathrm{BR}_1$ and $\mathrm{BR}_2$ refer to
$\mathrm{BR}(L\rightarrow Ze)$ and $\mathrm{BR}(Z\rightarrow
e^+e^-,~\mu^+\mu^-)$, respectively.  As shown in
Table~\ref{tab:table3} production of heavy leptons provides a
clean signature with small backgrounds. According to Poisson
statistics, the statistical significance (SS) of the signal for
the discovery at 95\% confidence level obeys the relation,
\begin{eqnarray}\label{e5}
\frac{\mathrm{S}}{\sqrt{\mathrm{S}+\mathrm{B}}}\geq 3,
\end{eqnarray}
where S and B are the numbers of signal and background events,
respectively. LHeC, with an integrated luminosity of $10^4
~\mathrm{pb}^{-1}$ yields 12 events per year for 800 GeV leptons,
hence a good place to search for heavy leptons. On the other hand
THERA seems capable for only about five 100 GeV leptons with
SS$\simeq 2$ which is not sufficient for observation.
\section{Single and Pair Production of Heavy Leptons in $e^+e^-$ Collisions}
\subsection{Single Production} 
\begin{figure*}
\includegraphics[width=6cm, angle=-90]{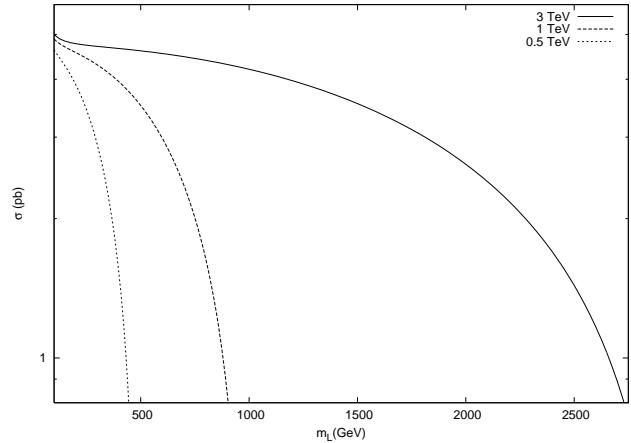}
\caption{The total cross sections as function of the heavy lepton
masses for linear colliders ILC ($\sqrt{s}=$0.5 TeV) and CLIC
($\sqrt{s}=$1 and 3 TeV) for single production.}\label{fig:f5}
\end{figure*}

\begin{figure*}
\includegraphics[width=6cm, angle=-90]{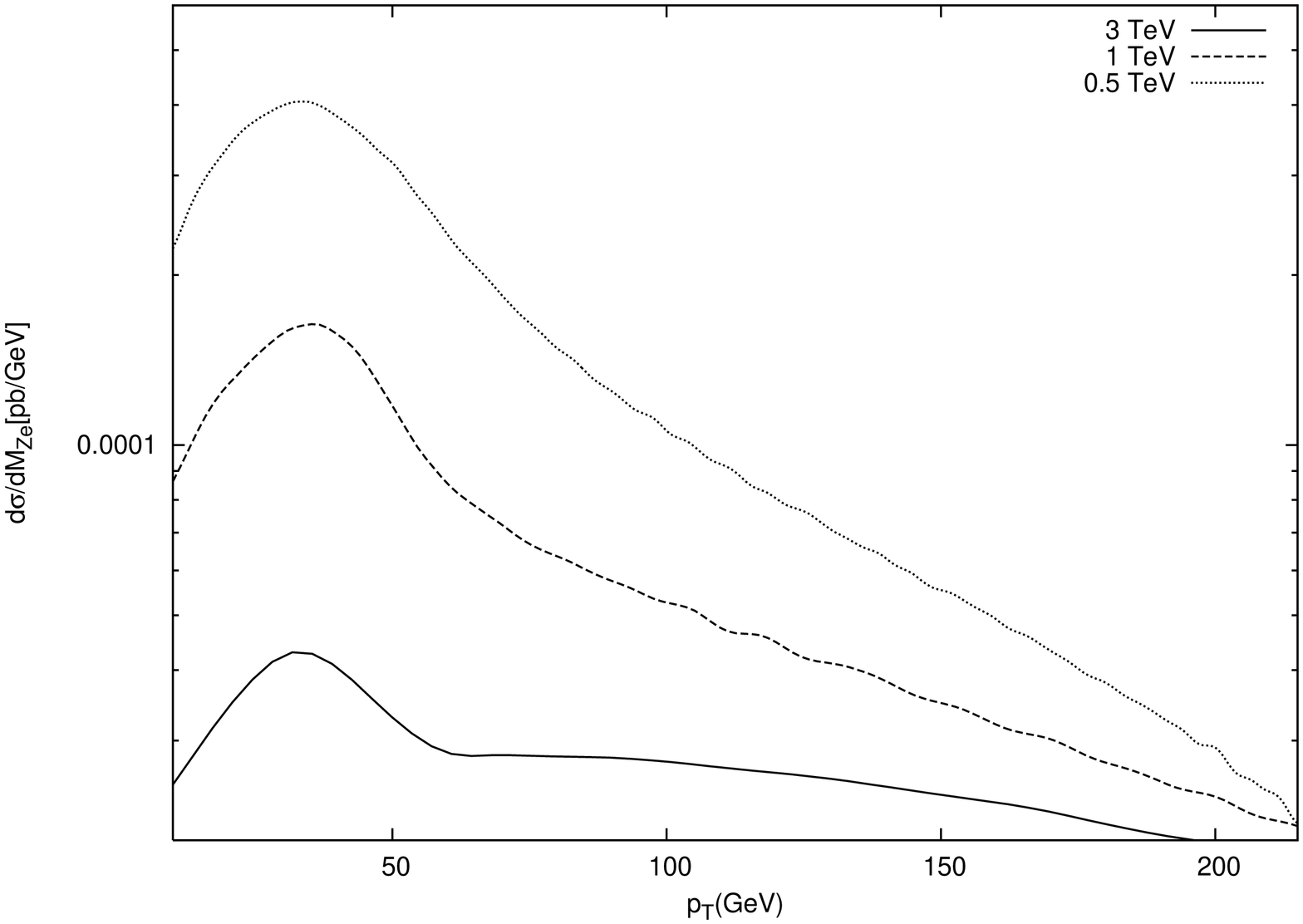}
\caption{$p_T$ distribution of the background for linear colliders
ILC and CLIC for single production.}\label{fig:f6}
\end{figure*}

\begin{figure*}
\includegraphics[width=6cm, angle=-90]{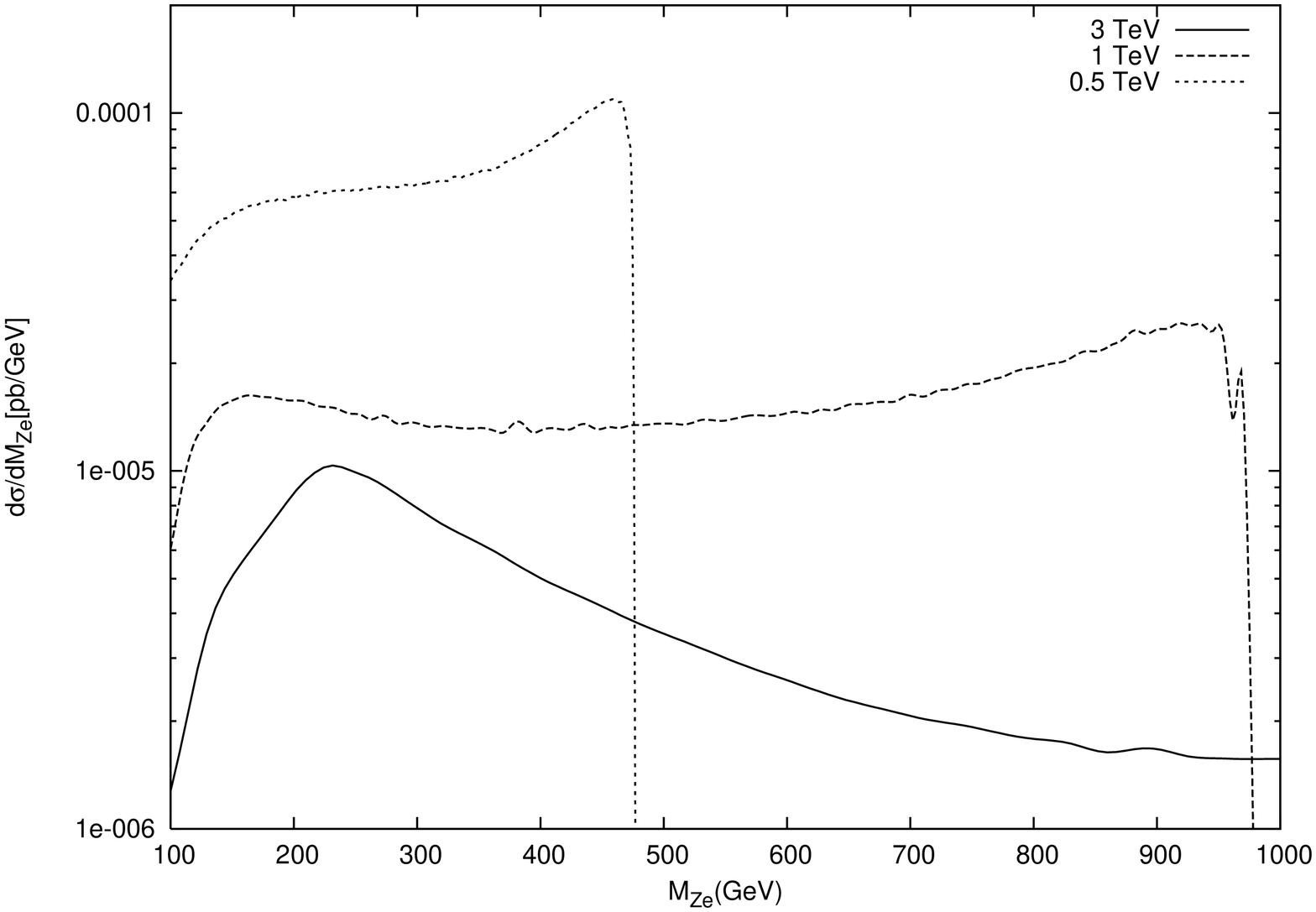}
\caption{The invariant mass distribution of the $Ze$ system for
the background for linear colliders ILC and CLIC for single
production.}\label{fig:c6_f5}
\end{figure*}

The single production of heavy
leptons $L$, in $e^-e^+$ collisions occur through the $s$ and $t$
channels via the process $e^-e^+\rightarrow L^-e^+$ represented by
four Feynman diagrams, two for $Z$ mediation plus two for photon
mediation. The corresponding total cross section is obtained as
follows;
\begin{eqnarray}\label{e66}
\frac{d\sigma}{dt}&=&\frac{\pi\alpha^2}{8\Lambda^2s^2\sin^4\theta_W
\cos^4\theta_W[(s-M_Z^2)^2+M_Z^2\Gamma_Z^2][(t-M_Z^2)^2
+M_Z^2\Gamma_Z^2]} \nonumber\\&&\times\bigg\{2\Lambda
\kappa_Zc_vst(t-s)m_L[c_a^2(7(s+t)-6m_L^2)+c_v^2(s+t-2m_L^2)]
-\kappa_Z^2(c_a^2+c_v^2)\nonumber\\&&\times
st[(s+t)m_L^4-2(s^2+t^2)m_L^2
+2(s^3+t^3)]+\Lambda^2[(c_a^4+c_v^4)(2(s^4-t^2s^2
+t^4)\nonumber\\&&-(s+t)(2s^2-3st+2t^2)m_L^2)-2c_a^2c_v^2(
(s-2t)(2s-t)(s+t)m_L^2\nonumber\\&&-2(s^4-3t^2s^2+t^4))]
+M_Z^2[2\kappa_Z^2(c_a^2+c_v^2)(2m^4-2(s+t)m_L^2+(s+t)^2)st
\nonumber\\&&+2\kappa_Zc_v\Lambda
(s-t)m_L[c_a^2(3s^2+3t^2+8st-3(s+t)m_L^2)+c_v^2
(s^2+t^2-(s+t)m_L^2)]\nonumber\\&&-2\Lambda^2(c_a-c_v)^2(c_a+c_v)^2(s^3+t^3
-(s^2+t^2)m_L^2)-(M_Z^2+\Gamma_Z^2)\nonumber\\&&\times
[\kappa_Z^2(c_a^2+c_v^2)(s+t)
(m_L^4-(s+t)m_L^2+2st)+\Lambda^2(c_a-c_v)^2(c_a+c_v)^2\nonumber\\&&
\times((s+t)m_L^2-s^2-t^2)]]\bigg\}
-\frac{2\pi\alpha^2\kappa_\gamma^2}
{\Lambda^2s^3t}\bigg\{(s+t)m_L^4
-2(s^2+t^2)m_L^2+2(s^3+t^3)\bigg\}\nonumber\\&&+\frac{\pi\alpha^2\kappa_\gamma
M_Z \Gamma_Z}{\Lambda^2s^2\sin^2\theta_W
\cos^2\theta_W[(t-M_Z^2)^2+M_Z^2\Gamma_Z^2]}
\bigg\{\kappa_Zc_vm_L^4-2\Lambda
(c_a^2+c_v^2)m_L^3\nonumber\\&&-2\kappa_Zc_vm_L^2s+\Lambda
(c_a^2(3s+2t)+c_v^2(s+2t))m_L
+\kappa_Zc_v(2s-t)(s+t)\bigg\}\nonumber\\&&+\frac{\pi\alpha^2\kappa_\gamma
M_Z \Gamma_Z}{\Lambda^2s^2\sin^2\theta_W
\cos^2\theta_W[(s-M_Z^2)^2+M_Z^2\Gamma_Z^2]}
\bigg\{\kappa_Zc_v(m_L^4-2tm_L^2-(s-2t)(s+t))\nonumber\\&&+\Lambda
((c_a^2+c_v^2)(2m_L^2-2s-t)-2c_a^2t)m_L\bigg\}
\end{eqnarray}
To compare the linear colliders, we display the total cross
sections as functions of the heavy lepton masses in
Fig.~\ref{fig:f5} by assuming $\Lambda=m_L$ and taking the
anomalous coupling parameters as $\kappa_\gamma=\kappa_Z=0.02$.
Here, we applied the same cut, $|M_{Ze}-m_L|<50$ GeV, as in $ep$
collisions. Fig.~\ref{fig:f6} shows the $p_T$ distributions of the
final state particles at three different linear colliders. The
character of this distribution is similar to that of $ep$
collisions. In Fig.~\ref{fig:c6_f5} we give the invariant mass
distributions $M_{Ze}$, with cut $p_T^{e,j}>50$ GeV.

In Tables~\ref{tab:table4}, \ref{tab:table5} and \ref{tab:table6},
we presented the signal and total background cross sections for
the single production of heavy lepton depending on its mass $m_L$,
for 0.5, 1 and 3 TeV energy $e^-e^+$ colliders, respectively. As
seen from these tables, the SS values are high enough for
observability limit, for mass values up to the center of mass
energies of the colliders.

In the single production case we expect 777 events per year for
450 GeV leptons, 903 events for 900 GeV leptons and 810 events for
2750 GeV leptons at 0.5, 1 and 3 TeV, respectively.
\subsection{Pair Production}
\begin{figure*}
\includegraphics[width=6cm, angle=-90]{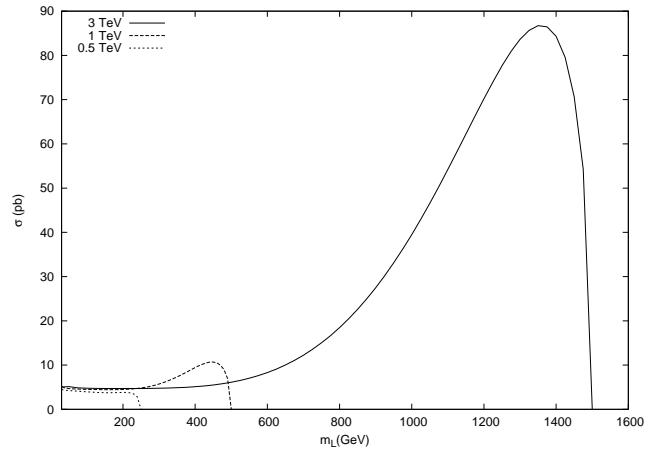}
\caption{The total cross sections as function of the heavy lepton
masses for linear colliders ILC and CLIC for pair
production.}\label{fig:c6_f10}
\end{figure*}

Pair production of heavy leptons via anomalous couplings occur
through the $t$-channel flavor changing neutral current process
$e^-e^+\rightarrow L^-L^+$. The differential cross section has the
form;
\begin{eqnarray}\label{e4_2}
\frac{d\sigma_{tot.}}{dt}&=&\frac{\pi\alpha^2\kappa_\gamma^2}{\Lambda^2s^2t^2}
\bigg\{2m_L^8-4tm_L^6+t(4s+3t)m_L^4-2t^2(4s+t)m_L^2+t^2(2s+t)^2\bigg\}\nonumber\\
&+&\frac{\pi\alpha^2}{16M_Z^4\Lambda^4s^2\sin^4\theta_W\cos^4\theta_W\left[
(t-M_Z^2)^2+M_Z^2\Gamma_Z^2\right]}\nonumber\\
&\times&\bigg\{\Lambda^4(c_a^2+c_v^2)m_L^4[(m_L^2-t)^2+4M_Z^2s]\nonumber\\
&-&2\Lambda^2M_Z^2
\kappa_Z^2c_v^2m_L^2t[m_L^4+t(t+2s-2m_L^2)]\nonumber\\
&+&M_Z^4[2\Lambda^4[(c_a^4+c_v^4)(m_L^4-2(2s+t)m_L^2+2s^2+t^2+2st)\nonumber\\
&+&2c_a^2c_v^2(3m_L^4-2(2s+3t)m_L^2+2s^2+3t^2+6st)]\nonumber\\
&-&2\kappa_Z^2\Lambda^2[4(c_a^2+c_v^2)s(m_L^4+st+t^2)+(c_a^2+2c_v^2)(m_L^6-2tm_L^4+t^2m_L^2)
\nonumber\\&-&10c_v^2tsm_L^2]+\kappa_Z^2(2m_L^8-4tm_L^6+tm_L^4(4s+3t)
\nonumber\\&-&2t^2(4s+t)m_L^2+t^2(2s+t)^2)]\bigg\}\nonumber\\
&+&\frac{\pi\alpha^2\kappa_\gamma^2(t-M_Z^2)}{2M_Z^2\Lambda^4s^2t\sin^2
\theta_W\cos^2\theta_W\left[(t-M_Z^2)^2+M_Z^2\Gamma_Z^2\right]}\nonumber\\
&\times&\bigg\{\kappa_Z^2M_Z^2[2m_L^8-4tm_L^6+t(4s+3t)m_L^4-2t^2(4s+t)m_L^2\nonumber\\
&+&t^2(2s+t)^2]-2\Lambda^2M_Z^2m_L^2[(c_a^2+c_v^2)(m_L^4+t^2-2tm_L^2)\nonumber\\
&+&st(2c_a^2-c_v^2)]-c_v^2\Lambda^2tm_L^2[m_L^4-2tm_L^2+(2s+t)t]\bigg\}
\end{eqnarray}
The total cross sections as functions of heavy lepton masses $m_L$
are displayed in Fig.~\ref{fig:c6_f10} for three options.

Signal and total background cross sections depending again on the
heavy lepton masses, are presented in Tables~\ref{tab:table7},
\ref{tab:table8} and \ref{tab:table9} at 0.5, 1 and 3 TeV,
respectively.

\begin{figure*}
\includegraphics[width=6cm, angle=-90]{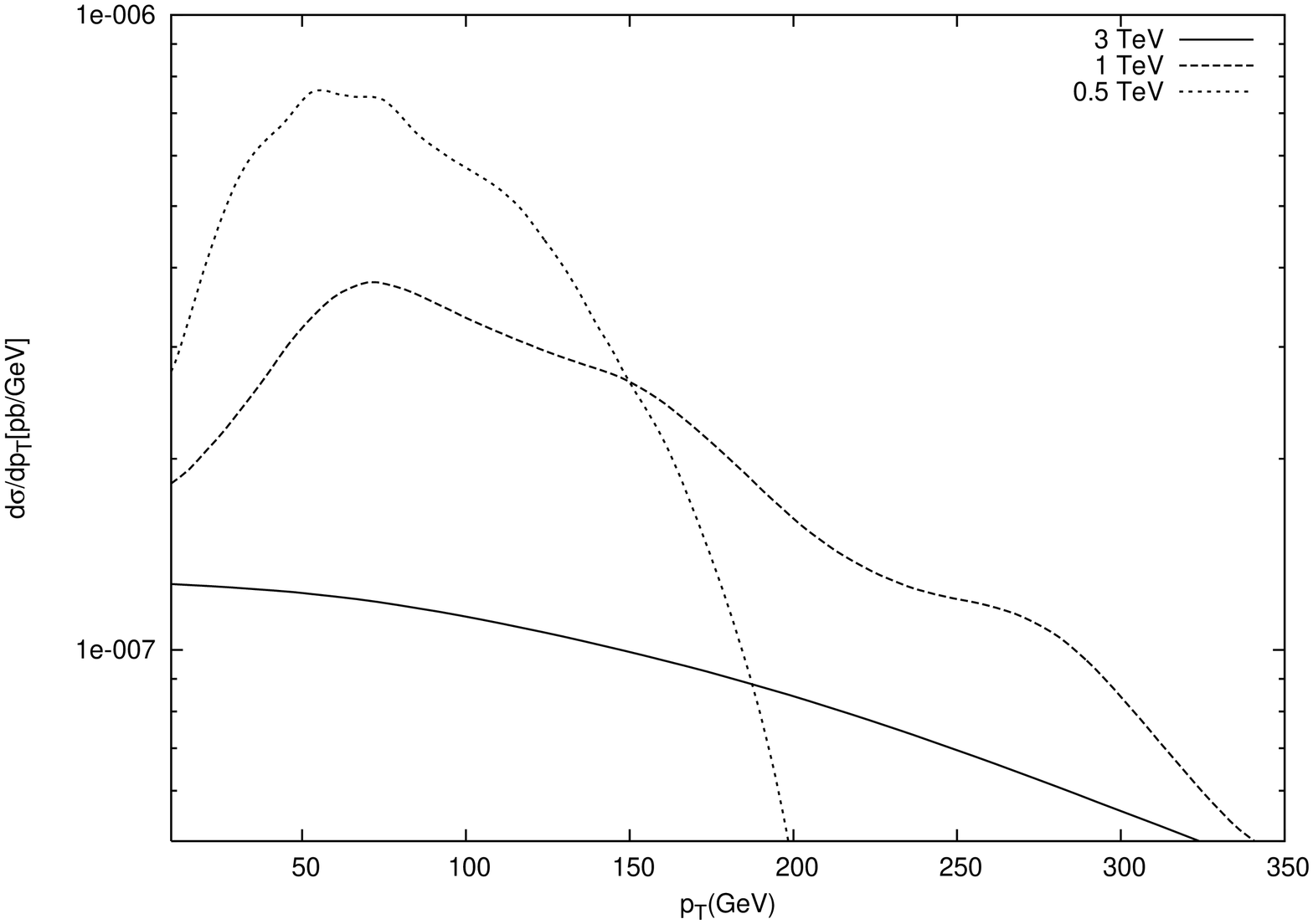}
\caption{$p_T$ distribution of the background for linear colliders
ILC and CLIC for pair production.}\label{fig:c6_f11}
\end{figure*}

\begin{figure*}
\includegraphics[width=6cm, angle=-90]{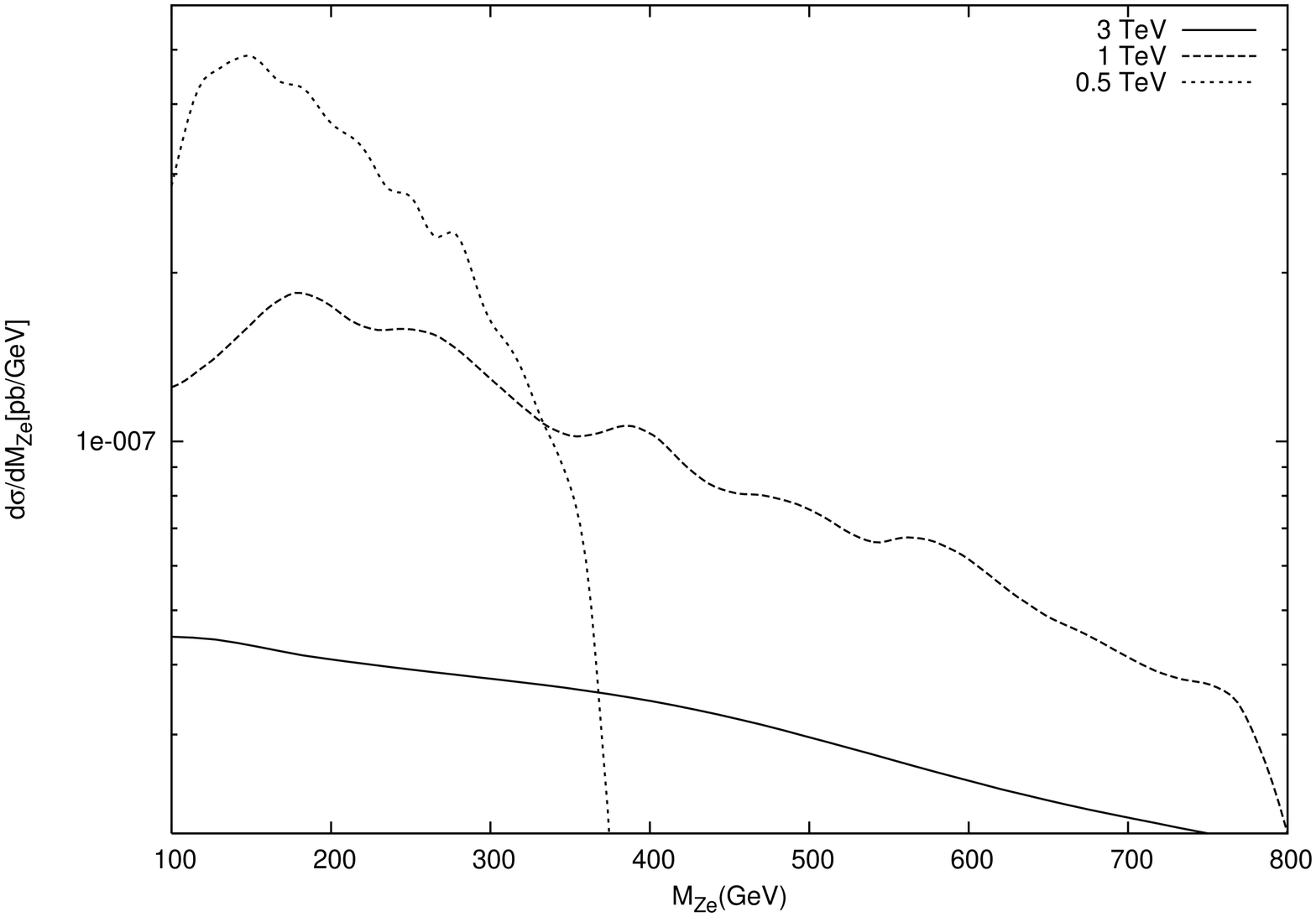}
\caption{The invariant mass distribution of the $Ze$ system for
the background for linear colliders ILC and CLIC for pair
production.}\label{fig:c6_f14}
\end{figure*}

We applied an initial cut on the electron and jet transverse
momentum $p_{T}^{e,j}>50$ GeV for the background analysis.
Fig.~\ref{fig:c6_f11} show the $p_T$ distributions of the final
state particles at the colliders. The distribution of invariant
mass $M_{Ze}$ is presented in Fig.~\ref{fig:c6_f14} at
$\sqrt{s}=$0.5, 1 and 3 TeV.

From Table~\ref{tab:table9} it is seen that the number of signal
events could be as high as $8\times10^{4}$ for 1250 GeV leptons at
$\sqrt{s}=3$ TeV.
\section{Production of Heavy Leptons at the CERN LHC}
In $pp$ collisions heavy lepton production occurs via parton level
subprocess $q\bar{q}\rightarrow Ll$, which are $\gamma$ and $Z$
exchanged $s$ channel FCNC reactions.

The form of the differential cross section is as follows;
\begin{eqnarray}\label{e4}
\frac{d\hat{\sigma}(q\bar{q}\rightarrow
Ll)}{d\hat{t}}&=&\frac{2e_q^2\kappa_\gamma^2\pi\alpha^2}{\Lambda^2\hat{s}^3}
\bigg\{(\hat{s}+2\hat{t})m_L^2-m_L^4-2\hat{t}(\hat{s}+\hat{t})\bigg\}\nonumber\\
&&+\frac{\pi\alpha^2}{8\Lambda^2\hat{s}^2\sin^4\theta_W\cos^4\theta_W
[(\hat{s}-M_Z^2)^2+M_Z^2\Gamma_Z^2]}\nonumber\\
&&\times\bigg\{\kappa_Z^2({c_a^q}^2+{c_v^q}^2)\hat{s}((\hat{s}+2\hat{t})m_L^2
-2\hat{t}(\hat{s}+\hat{t})-m_L^4)\nonumber\\
&&+2\kappa_Z\Lambda
\hat{s}m_L(c_v({c_a^q}^2+{c_v^q}^2)(\hat{s}-m_L^2)-2c_ac_a^qc_v^q
(m_L^2-\hat{s}-2\hat{t}))\nonumber\\
&&+\Lambda^2[(c_a^2+c_v^2)({c_a^q}^2+{c_v^q}^2)
(\hat{s}^2+2\hat{t}^2+2\hat{s}\hat{t}-(\hat{s}+2\hat{t})m_L^2)
\nonumber\\
&&+4c_ac_vc_a^qc_v^q\hat{s}(\hat{s}+2\hat{t}-m_L^2)]
 \bigg\}\nonumber\\
&&+\frac{e_q\pi\alpha^2\kappa_\gamma
M_Z\Gamma_Z}{\Lambda^2\hat{s}^2\sin^2\theta_W\cos^2\theta_W
[(\hat{s}-M_Z^2)^2+M_Z^2\Gamma_Z^2]}\nonumber\\
&&\times
\bigg\{\kappa_Zc_v^q(m_L^4-(\hat{s}+2\hat{t})m_L^2+2\hat{t}
(\hat{s}+\hat{t}))\nonumber\\
&&+\Lambda
m_L(c_ac_a^q(m_L^2-\hat{s}-2\hat{t})+c_vc_v^q(m_L^2-\hat{s}))\bigg\}
\end{eqnarray}.

The total production cross section is obtained by the integration
over the parton distributions as;
\begin{eqnarray}\label{e7}
\sigma(q\bar{q}\rightarrow
Ll)=\int_{m_L^2/s}^{1}d\tau\int_{\tau}^{1}\frac{dx}{x}[f_{q/p}(x,Q^2)f_{\bar{q}}(x/\tau,Q^2)+
f_{q}(x/\tau,Q^2)f_{\bar{q}}(x,Q^2)] \hat{\sigma}(\hat{s})
\end{eqnarray}
where $\hat{s}=\tau s$. As in the $ep$ collisions, here we use the
CTEQ5 parametrization with $Q=m_L$ and we take
$\kappa_\gamma=\kappa_Z=0.02$ and $\Lambda=m_L$.

\begin{figure*}
\includegraphics[width=6cm, angle=-90]{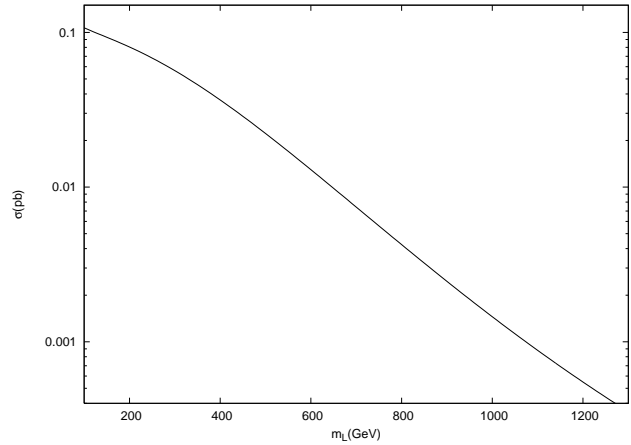}
\caption{The total cross sections as function of the heavy lepton
masses for LHC ($\sqrt{s}=$14 TeV).}\label{fig:c7_f1}
\end{figure*}

\begin{figure*}
\includegraphics[width=6cm, angle=-90]{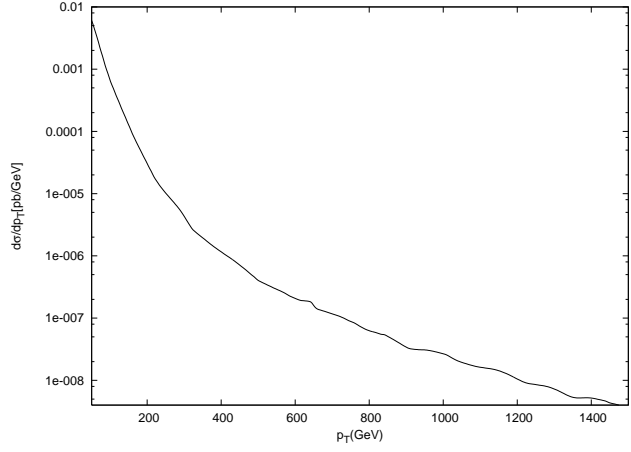}
\caption{$p_T$ distribution of the background for
LHC.}\label{fig:c7_f2}
\end{figure*}

\begin{figure*}
\includegraphics[width=6cm, angle=-90]{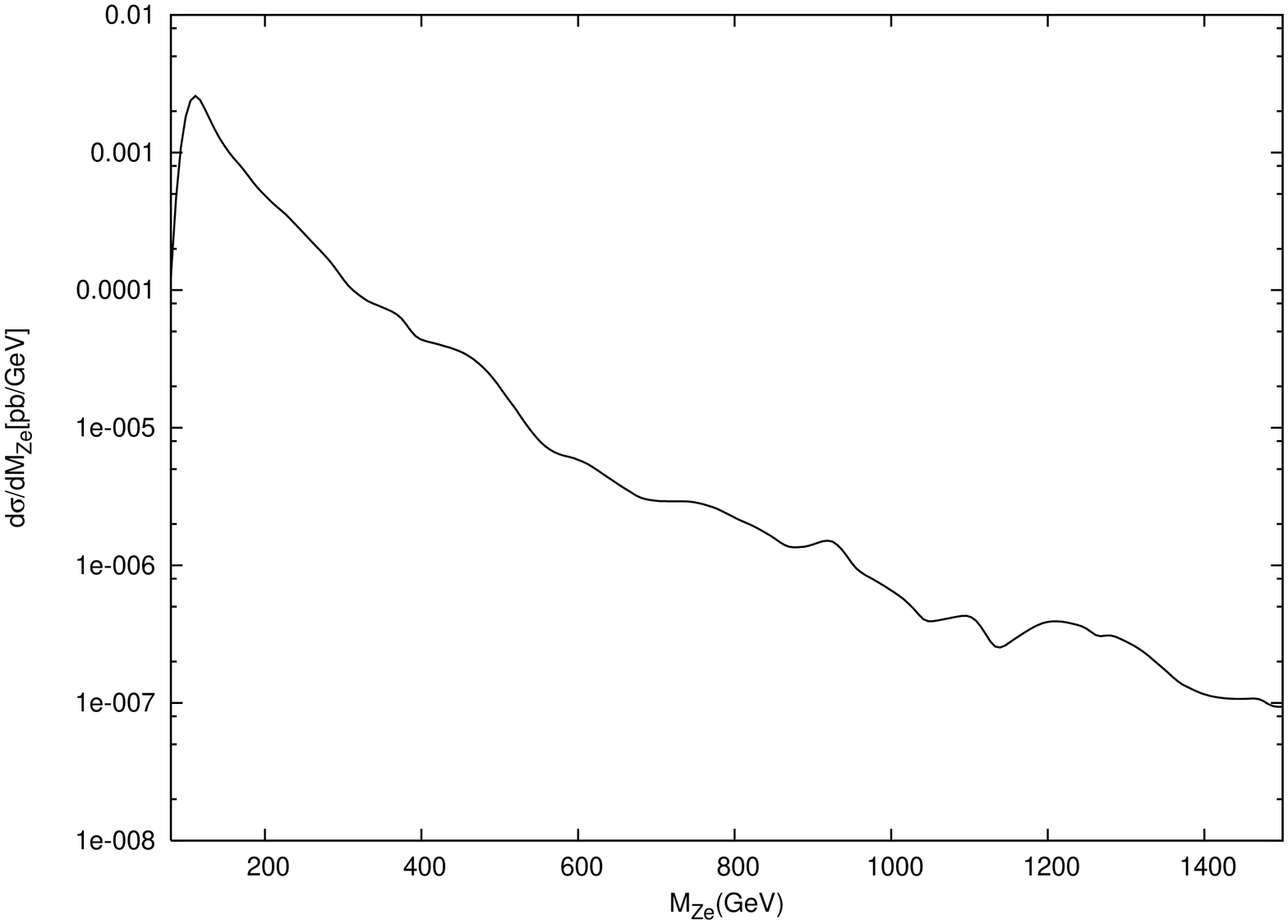}
\caption{The invariant mass distribution of the $Ze$ system for
the background for LHC.}\label{fig:c7_f3}
\end{figure*}

We give the production cross sections for the signal as functions
of the heavy lepton masses, $m_L$, in Fig.~\ref{fig:c7_f1} for the
center of mass energy of $\sqrt{s}=14$ TeV. We applied an optimal
cut of $|M_{Ze}-m_L|<50$ GeV for the considered heavy lepton mass
range. Fig.~\ref{fig:c7_f2} shows the $p_T$ distribution of the
final state particles for the relevant background process
$q\bar{q}\rightarrow eZl$ in $pp$ collisions. This $p_T$
distribution has a peak around 50 GeV and suppressed at higher
values. Fig.~\ref{fig:c7_f3} shows the invariant mass
distributions of $Ze$ system, after a cut of $p_T^{e,j}>50$ GeV.

In Table~\ref{tab:table10}, we present the production cross
sections and the number of signal events depending on the heavy
lepton masses for a suitable range. The branching ratios
$\mathrm{BR}_1$ and $\mathrm{BR}_2$ refer to
$\mathrm{BR}(L\rightarrow Ze)$ and $\mathrm{BR}(Z\rightarrow
e^+e^-,~\mu^+\mu^-)$, respectively. Production of heavy leptons
provides a clean signature, as shown in this table. Taking into
account the SS, LHC can observe heavy leptons with masses up to
about 650 GeV. For 650 GeV leptons we expect 11 events per year at
the LHC.

We have used the high energy package CompHEP for calculations of
background cross sections reported in this study
\cite{Pukhov:1999gg}.
\section{conclusion}
In conclusion, the present work gives an analysis of possible
production of heavy leptons via anomalous interactions in $ep$,
$e^+e^-$ and $pp$ collisions. It is shown that, after kinematical
cuts a statistical significance of $\mathrm{SS}\geq 3$ can be
achieved and for anomalous magnetic moment factors of
$\kappa_\gamma=\kappa_Z=0.02$, heavy leptons with masses about 800
GeV at LHeC, 450 GeV at ILC and 650 GeV at the LHC can be
observed. Among the others ILC seems the most convenient place to
search for heavy leptons as providing a high event number of
$8\times 10^4$. LHeC and LHC are also good places for searching
heavy leptons besides the other major high energy experiments.
Hence, these collider options seem to be capable of probing new
physics in the case of anomalous interactions are valid as being
an underlying theory.

\begin{acknowledgments} This study was partially supported by Abant
Izzet Baysal University Research fund (2005.03.02.216).
\end{acknowledgments}


\newpage
\begin{table}
\caption{Main parameters of $ep$ and $e^+e^-$ and $pp$ colliders,
$L^{int}$ denotes the integrated luminosity for one
year.}\label{tab:table1}
\begin{tabular}{lcccc} \hline \hline
$ep$ Colliders & $E_e$ (TeV) & $E_p$ (TeV) & $\sqrt{s_{ep}}$ (TeV) & $L^{int}_{ep}(\mathrm{pb}^{-1})$  \\
  \hline
   THERA & 0.25 & 1 & 1 & 40 \\
   LHeC& 0.07 & 7 & 1.4 & $ 10^{4}$\\
   \hline\hline
$e^+e^-$ Colliders & $E_{e^+}$ (TeV) & $E_{e^-}$ (TeV) & $\sqrt{s_{e^+e^-}}$ (TeV) & $L^{int}_{e^+e^-}(\mathrm{pb}^{-1})$  \\
  \hline
   ILC  &0.25&0.25& 0.5& $10^{5}$\\
CLIC &0.5&0.5&1.0 & $10^{5}$\\
CLIC &1.5&1.5&3.0 & $10^{5}$\\
\hline\hline
$pp$ Collider & $E_p$ (TeV) & $E_p$ (TeV) & $\sqrt{s_{pp}}$ (TeV) & $L^{int}_{pp}(\mathrm{pb}^{-1})$  \\
  \hline
  LHC& 7 & 7 & 14 & $ 10^{4}$\\ \hline \hline
\end{tabular}
\end{table}

\begin{table}
\caption{Decay widths of heavy leptons.}\label{tab:table2}
\begin{tabular}{cccc} \hline \hline
 $m_L$ (GeV) & $\Gamma(L\rightarrow\gamma e)$(GeV) & $\Gamma(L\rightarrow Ze)$(GeV) &  $\Gamma_{\mathrm{Total}}$ (GeV)  \\
  \hline
   250& $7.30\times10^{-4}$ & 3   & 3   \\
   500& $1.46\times10^{-3}$ & 21  & 21  \\
   750& $2.19\times10^{-3}$ & 72  & 72  \\
  1000& $2.92\times10^{-3}$ & 171 & 171 \\
  1250& $3.65\times10^{-3}$ & 333 & 333 \\
  1500& $4.38\times10^{-3}$ & 576 & 576 \\
  1750& $5.11\times10^{-3}$ & 914 & 914 \\
  2000& $5.84\times10^{-3}$ & 1365& 1365\\
  2250& $6.57\times10^{-3}$ & 1943& 1943\\
  2500& $7.30\times10^{-3}$ & 2666& 2666\\ \hline \hline
\end{tabular}
\end{table}

\begin{table}
\caption{The signal and background cross sections and SS depending
on the heavy lepton masses for THERA with $\sqrt{s}=1$
TeV.}\label{tab:tab3}
\begin{tabular}{ccccccc} \hline \hline
   $m_L$ (GeV) &  $\sigma$ (pb) &  $\sigma \times \mathrm{BR}_1$ (pb)& $\sigma \times \mathrm{BR}_1 \times \mathrm{BR}_2$ (pb) &  $\sigma_B$(pb)  & $S/\sqrt{S+B}$  \\
  \hline
   100 & 12.26& 4.05 & 0.13 & $4.53\times10^{-5}$ & 2 \\
   150 & 8.21 & 2.71 & 0.09 & $2.54\times10^{-4}$ & 2 \\
   200 & 5.50 & 1.82 & 0.06 & $2.39\times10^{-4}$ & 2 \\ \hline \hline
\end{tabular}
\end{table}

\begin{table}
\caption{The signal and background cross sections and SS depending
on the heavy lepton masses for LHeC with $\sqrt{s}=1.4$
TeV.}\label{tab:table3}
\begin{tabular}{ccccccc} \hline \hline
   $m_L$ (GeV) &  $\sigma$ (pb) &  $\sigma \times \mathrm{BR}_1$ (pb)& $\sigma \times \mathrm{BR}_1 \times \mathrm{BR}_2$ (pb) &  $\sigma_B$(pb)  & $S/\sqrt{S+B}$  \\
  \hline
   200 & 9.43 & 3.11 & 0.194 & $4.75\times10^{-4}$ & 32 \\
   400 & 2.80 & 0.93 & 0.031 & $2.28\times10^{-4}$ & 17 \\
   600 & 0.68 & 0.22 & 0.007 & $8.69\times10^{-5}$ & 9 \\
   800 & 0.11 & 0.04 & 0.001 & $2.19\times10^{-5}$ & 3 \\ \hline \hline
\end{tabular}
\end{table}

\begin{table}
\begin{center}\caption{The signal and background cross
sections and SS depending on the heavy lepton masses at
$\sqrt{s}=0.5$ TeV for single production.}\label{tab:table4}
\begin{tabular}{cccccc}\hline\hline
  $m_L$ (GeV) &  $\sigma$ (pb) &  $\sigma \times \mathrm{BR}_1$ (pb)& $\sigma \times \mathrm{BR}_1 \times \mathrm{BR}_2$ (pb) &  $\sigma_B$(pb)  & $S/\sqrt{S+B}$  \\
  \hline
   100& 4.63 & 1.53 & 0.0504 &  $1.28\times10^{-2}$ & 63 \\
   200& 3.90 & 1.29 & 0.0424 &  $5.10\times10^{-3}$ & 62  \\
   300& 2.88 & 0.95 & 0.0314 &  $2.05\times10^{-3}$ & 54  \\
   400& 1.51 & 0.50 & 0.0165 &  $4.83\times10^{-4}$ & 40 \\\hline\hline
\end{tabular}
\end{center}
\end{table}

\begin{table}
\begin{center}\caption{The signal and background cross
sections and SS depending on the heavy lepton masses at
$\sqrt{s}=1$ TeV for single production.}\label{tab:table5}
\begin{tabular}{cccccc}\hline\hline
  $m_L$ (GeV) &  $\sigma$ (pb) &  $\sigma \times \mathrm{BR}_1$ (pb)& $\sigma \times \mathrm{BR}_1 \times \mathrm{BR}_2$ (pb) &  $\sigma_B$(pb)  & $S/\sqrt{S+B}$  \\
  \hline
   100& 4.89 & 1.61 & 0.053 &  $1.43\times10^{-2}$ & 65  \\
   300& 4.30 & 1.42 & 0.047 &  $4.60\times10^{-3}$ & 65  \\
   500& 3.51 & 1.16 & 0.038 &  $2.36\times10^{-3}$ & 60  \\
   700& 2.35 & 0.78 & 0.026 &  $1.04\times10^{-3}$ & 50  \\
   900& 0.83 & 0.27 & 0.009 &  $1.81\times10^{-4}$ & 30  \\\hline\hline
\end{tabular}
\end{center}
\end{table}

\begin{table}
\begin{center}
\caption{The signal and background cross sections and SS depending
on the heavy lepton masses at $\sqrt{s}=3$ TeV for single
production.}\label{tab:table6}
\begin{tabular}{ccccccc}\hline\hline
   $m_L$ (GeV) &  $\sigma$ (pb) &  $\sigma \times \mathrm{BR}_1$ (pb)& $\sigma \times \mathrm{BR}_1 \times \mathrm{BR}_2$ (pb) &  $\sigma_B$(pb)  & $S/\sqrt{S+B}$  \\
  \hline
   250 & 4.74 & 1.56 & 0.052  & $6.37\times10^{-3}$ & 68 \\
   750 & 4.44 & 1.46 & 0.048  & $3.55\times10^{-3}$ & 67 \\
   1250& 3.91 & 1.29 & 0.043  & $3.04\times10^{-3}$ & 63 \\
   1750& 3.12 & 1.03 & 0.034  & $2.44\times10^{-3}$ & 56 \\
   2250& 2.06 & 0.68 & 0.022  & $1.43\times10^{-3}$ & 46 \\
   2750& 0.74 & 0.25 & 0.008  & $3.06\times10^{-4}$ & 28 \\
   \hline\hline
\end{tabular}
\end{center}
\end{table}

\begin{table}
\begin{center}
\caption{The signal and background cross sections and SS depending
on the heavy lepton masses at $\sqrt{s}=0.5$ TeV for pair
production.}\label{tab:table7}
\begin{tabular}{cccccc}\hline\hline
  $m_L$ (GeV) &  $\sigma$ (pb) &  $\sigma \times \mathrm{BR}_1$ (pb)& $\sigma \times \mathrm{BR}_1 \times \mathrm{BR}_2$ (pb) &  $\sigma_B$(pb)  & $S/\sqrt{S+B}$  \\
  \hline
   100& 3.95 & 1.30 & 0.0430 &  $3.22\times10^{-2}$ & 50 \\
   150& 3.76 & 1.24 & 0.0409 &  $9.20\times10^{-3}$ & 58  \\
   200& 3.83 & 1.26 & 0.0417 &  $1.78\times10^{-3}$ & 63  \\
   240& 2.79 & 0.92 & 0.0304 &  $1.98\times10^{-2}$ & 43  \\\hline\hline
\end{tabular}
\end{center}
\end{table}

\begin{table}
\begin{center}
\caption{The signal and background cross sections and SS depending
on the heavy lepton masses at $\sqrt{s}=1$ TeV for pair
production.}\label{tab:table8}
\begin{tabular}{cccccc}\hline\hline
  $m_L$ (GeV) &  $\sigma$ (pb) &  $\sigma \times \mathrm{BR}_1$ (pb)& $\sigma \times \mathrm{BR}_1 \times \mathrm{BR}_2$ (pb) &  $\sigma_B$(pb)  & $S/\sqrt{S+B}$  \\
  \hline
   100& 4.51 & 1.49 & 0.0491 &$9.92\times10^{-3}$ & 64  \\
   200& 4.48 & 1.48 & 0.0488 &$4.91\times10^{-3}$ & 67  \\
   300& 5.72 & 1.89 & 0.0623 &$7.31\times10^{-3}$ & 75  \\
   400& 9.44 & 3.12 & 0.1028 &$4.24\times10^{-3}$ & 99  \\\hline\hline
\end{tabular}
\end{center}
\end{table}

\begin{table}
\begin{center}
\caption{The signal and background cross sections and SS depending
on the heavy lepton masses at $\sqrt{s}=3$ TeV for pair
production.}\label{tab:table9}
\begin{tabular}{ccccccc}\hline\hline
  $m_L$ (GeV) &  $\sigma$ (pb) &  $\sigma \times \mathrm{BR}_1$ (pb)& $\sigma \times \mathrm{BR}_1 \times \mathrm{BR}_2$ (pb) &  $\sigma_B$(pb)  & $S/\sqrt{S+B}$  \\
  \hline
   250 &  4.72 &  1.56 & 0.0514 &  $2.11\times10^{-3}$ & 70  \\
   500 &  6.12 &  2.02 & 0.0666 &  $3.32\times10^{-4}$ & 81  \\
   750 & 15.03 &  4.96 & 0.1637 &  $4.11\times10^{-4}$ & 128 \\
   1000& 39.49 & 13.03 & 0.4300 &  $1.93\times10^{-4}$ & 207 \\
   1250& 77.64 & 25.62 & 0.8455 &  $1.32\times10^{-5}$ & 291 \\
   \hline\hline
\end{tabular}
\end{center}
\end{table}
\clearpage
\begin{table}
\begin{center}
\caption{The signal and background cross sections and SS depending
on the heavy lepton masses for the LHC ($\sqrt{s}=14$
TeV).}\label{tab:table10}
\begin{tabular}{ccccccc}\hline\hline
  $m_L$ &  $\sigma\times 10^{-2}$  &
  $\sigma \times \mathrm{BR}_1\times 10^{-2}$&
  $\sigma \times \mathrm{BR}_1 \times \mathrm{BR}_2\times 10^{-4}$&
  $\sigma_B\times 10^{-4}$  & $S/\sqrt{S+B}$  \\
(GeV) & (pb) & (pb)& (pb) & (pb) &   &   \\
  \hline
   150 & 9.22 & 3.04 & 10.0&  $4.96$ & 8 \\
   250 & 6.80 & 2.24 &  7.4&  $4.05$ & 7 \\
   350 & 4.58 & 1.51 &  5.0&  $2.80$ & 6 \\
   450 & 2.86 & 0.95 &  3.1&  $1.84$ & 4 \\
   550 & 1.70 & 0.56 &  1.9&  $1.14$ & 3 \\
   650 & 0.98 & 0.32 &  1.1&  $0.65$ & 3 \\\hline\hline
\end{tabular}
\end{center}
\end{table}

\newpage

\begin{thebibliography}{99}
\bibitem{Almeida:1990ay}
  F.~M.~L.~Almeida, J.~H.~Lopes, J.~A.~Martins Simoes and C.~M.~Porto,
  Phys.\ Rev.\ D {\bf 44}, 2836 (1991).
\bibitem{Almeida:1994ad}
  F.~M.~L.~Almeida, J.~H.~Lopes, J.~A.~Martins Simoes, P.~P.~Queiroz Filho and A.~J.~Ramalho,
  Phys.\ Rev.\ D {\bf 51}, 5990 (1995).
\bibitem{Almeida:2000yx}
  F.~M.~L.~Almeida, Y.~A.~Coutinho, J.~A.~Martins Simoes and M.~A.~B.~do Vale,
  Phys.\ Rev.\ D {\bf 63}, 075005 (2001).
\bibitem{Almeida:2003cy}
  F.~M.~L.~Almeida, Y.~A.~Coutinho, J.~A.~Martins Simoes, S.~Wulck and M.~A.~B.~do Vale,
  Eur.\ Phys.\ J.\ C {\bf 30}, 327 (2003).
\bibitem{Frampton:1992ik}
  P.~H.~Frampton, D.~Ng, M.~Sher and Y.~Yuan,
  Phys.\ Rev.\ D {\bf 48}, 3128 (1993).
\bibitem{Coutinho:1998bu}
  Y.~A.~Coutinho, J.~A.~Martins Simoes, C.~M.~Porto and P.~P.~Queiroz Filho,
  Phys.\ Rev.\ D {\bf 57}, 6975 (1998).
\bibitem{CiezaMontalvo:2002gk}
  J.~E.~Cieza Montalvo and P.~P.~de Queiroz Filho,
  Phys.\ Rev.\ D {\bf 66}, 055003 (2002).
\bibitem{CiezaMontalvo:2000ys}
  J.~E.~Cieza Montalvo and M.~D.~Tonasse,
  Nucl.\ Phys.\ B {\bf 623}, 325 (2002).
\bibitem{Almeida:1990vt}
  F.~M.~L.~Almeida, J.~A.~Martins Simoes and A.~J.~Ramalho,
  Nucl.\ Phys.\ B {\bf 347} (1990) 537.
\bibitem{Rizzo:1986wf}
  T.~G.~Rizzo,
  Phys.\ Lett.\ B {\bf 188}, 95 (1987).
\bibitem{Alan:2004rv}
  A.~T.~Alan, A.~T.~Tasci and O.~Cakir,
  Acta Phys.\ Polon.\ B {\bf 35}, 2199 (2004).
\bibitem{Alan:2006ux}
  A.~T.~Alan, A.~T.~Tasci and N.~Karagoz,
  Mod.\ Phys.\ Lett.\ A {\bf 21}, 1869 (2006).
\bibitem{Akrawy:1990dx}
  M.~Z.~Akrawy {\it et al.}  [OPAL Collaboration],
  Phys.\ Lett.\ B {\bf 240}, 250 (1990).
\bibitem{Decamp:1991uy}
  D.~Decamp {\it et al.}  [ALEPH Collaboration],
  Phys.\ Rept.\  {\bf 216}, 253 (1992).
\bibitem{Ahmed:1994yi}
  T.~Ahmed {\it et al.}  [H1 Collaboration],
  Phys.\ Lett.\ B {\bf 340}, 205 (1994).
\bibitem{Abramowicz:DESY-01-011}
  H.~Abramowicz {\it et al.} [TESLA-N Study Group], DESY-01-011.
\bibitem{Dainton:2006wd}
  J.~B.~Dainton, M.~Klein, P.~Newman, E.~Perez and F.~Willeke,
  DESY-06-006,
  arXiv:hep-ex/0603016.
\bibitem{lc}
International Linear Collider-Technical Review Report,
ILC-TRC/2003 Report (2003).
\bibitem{Accomando:2004sz}
  E.~Accomando {\it et al.}  [CLIC Physics Working Group],
  ``Physics at the CLIC multi-TeV linear collider''.
\bibitem{unknown:1995mi}
   D. Boussard et al., [LHC Study Group],
  CERN-AC-95-05-LHC (1995).
\bibitem{Lai:1999wy}
  H.~L.~Lai {\it et al.}  [CTEQ Collaboration],
  Eur.\ Phys.\ J.\ C {\bf 12}, 375 (2000).
\bibitem{Pukhov:1999gg}
  A.~Pukhov {\it et al.},
  arXiv:hep-ph/9908288.
\end{thebibliography}
\end{document}